\documentclass{emulateapj}

\usepackage{apjfonts}

\slugcomment{Draft}

\begin{document}

\shorttitle{First Star Formation with Relative Streaming}
\shortauthors{A. Stacy, V. Bromm and A. Loeb}

\title{Effect of Streaming Motion of Baryons Relative to Dark Matter
on the Formation of the First Stars}

\author{Athena Stacy$^{1}$\thanks{E-mail: minerva@astro.as.utexas.edu}, Volker Bromm$^{1}$ and Abraham Loeb$^{2}$}
\affil{$^{1}$Department of Astronomy and Texas Cosmology Center, The University of Texas at Austin, TX 78712\\
        $^{2}$Astronomy Department, Harvard University, 60 Garden Street, Cambridge, MA 02138, USA}

\begin{abstract}
We evaluate the effect of a supersonic relative velocity between the
baryons and dark matter on the thermal and density evolution of the
first gas clouds at $z \lesssim 50$.  Through a series of cosmological
simulations, initialized at $z_{\rm i}=100$ with a range of relative streaming
velocities and minihalo formation redshifts, we find that the typical
streaming velocities will have little effect on the gas evolution.
Once the collapse begins, the subsequent evolution of the gas will be
nearly indistinguishable from the case of no streaming, and star
formation will still proceed in the same way, with no change in the
characteristic Pop III stellar masses. Reionization is expected to be
dominated by halo masses of $\ga 10^8 M_\odot$, for which the
effect of streaming should be negligible.

\end{abstract}


\section{Introduction}
The formation of the first stars was a key event in the evolution of
the early universe
(e.g. \citealt{barkana&loeb2001,bromm&larson2004,ciardi&ferrara2005,glover2005,byhm2009,loeb2010}).
After the emission of the Cosmic Microwave Background at $z \sim
1000$, the universe entered the `Dark Ages,' the period when the
distribution of matter was very uniform and no luminous objects had
yet formed.  During this time, cold dark matter (DM) density
perturbations grew to make the halos inside of which the first stars
formed at $z \lesssim 50$.  These stars are believed to have formed
within $M\sim 10^6$\,M$_{\odot}$ minihalos, where the infall of the
baryons into the gravitational potential well of the DM-dominated
minihalo heated the gas sufficiently to enable H$_2$-driven cooling
and fragmentation
(e.g. \citealt{haimanetal1996,tegmarketal1997,yahs2003}).

The initial growth of the density fluctuations after recombination can
be described using linear perturbation theory, which assumes that
overdensities and velocity fields are small quantities.  Similarly,
cosmological simulations are initialized at high $z$ with small gas
and DM peculiar velocities, determined through a combination of the
$\Lambda$CDM model and Zeldovich approximation
(\citealt{zeldovich1970}).  Recently, \cite{tse&hirata2010} added a
complicating aspect to this picture by showing that at high redshift,
there is a supersonic relative velocity between the baryons and DM.
Whereas prior to recombination, photons and baryons are coupled such
that the baryonic sound speed is $\sim c/\sqrt{3}$, after
recombination the sound speed drops to $\sim 6$ km s$^{-1}$.  The
root-mean square relative velocity, on the other hand, is much higher,
30 km s$^{-1}$.  The relative velocities are dominated by modes on the
comoving scale of $\sim 150$ Mpc, the length scale of the sound
horizon at recombination, and are coherent on smaller scales of a few
Mpc.

\cite{tse&hirata2010} examined how this effect alters the growth of DM
structure, causing a small ($\sim 10\%$) suppression of the matter
power spectrum for modes with wavenumber $k \simeq 200 \rm Mpc^{-1}$.
Using the Press-Schechter formalism they have also found a decrease in
$M\sim 10^6$\,M$_{\odot}$ minihalos at high redshifts, $z = 40$.
They furthermore find that the relative velocity effect yields a
scale-dependent bias of the first halos. Extending upon this,
\cite{dalal&pen2010} analytically studied the impact of the relative
velocity on baryonic objects, finding that the collapse fraction will
be slightly reduced and that the large-scale clustering of $M\la 10^6$\,M$_{\odot}$ 
minihalos will be modulated on scales of $\sim 100$
Mpc.  The same applies to any observable 
that traces minihalos, including the 21 cm absorption power spectrum. 
 \cite{tseetal2010} find similar results in a more detailed analysis.

While these previous studies examined the large-scale effects of the
relative velocity, its direct influence on the delay of collapse and
the evolution of gas falling into a single minihalo has yet to be
considered.  Simulations are necessary to understand how the relative
streaming affects the non-linear regime and alters the processes
involved in the collapse of minihalo gas.  To this end, we perform a
set of cosmological simulations which include these streaming motions.
After the completion of this work, we became aware of an analogous
paper by \cite{maioetal2010}.  Similar to \cite{maioetal2010}, we find
a delay of gas collapse in early low-mass $M\sim 10^5-10^6$\,M$_{\odot}$ 
minihalos, but conclude that for typical
streaming velocities this delay will be negligible by $z\sim10$.  Our
work is complementary to that of \cite{maioetal2010} in that, while
they are able to find a $1-20\%$ overall suppression of the first
objects, our factor of $\sim$ 10 greater mass resolution allows us to
see the subsequent collapse of the gas to high densities, revealing
that even with relative streaming motions the thermal evolution of
primordial gas and subsequent Pop III star formation will be very
similar to no-streaming cases following the initial collapse.

\begin{figure*}
\includegraphics[width=.7\textwidth]{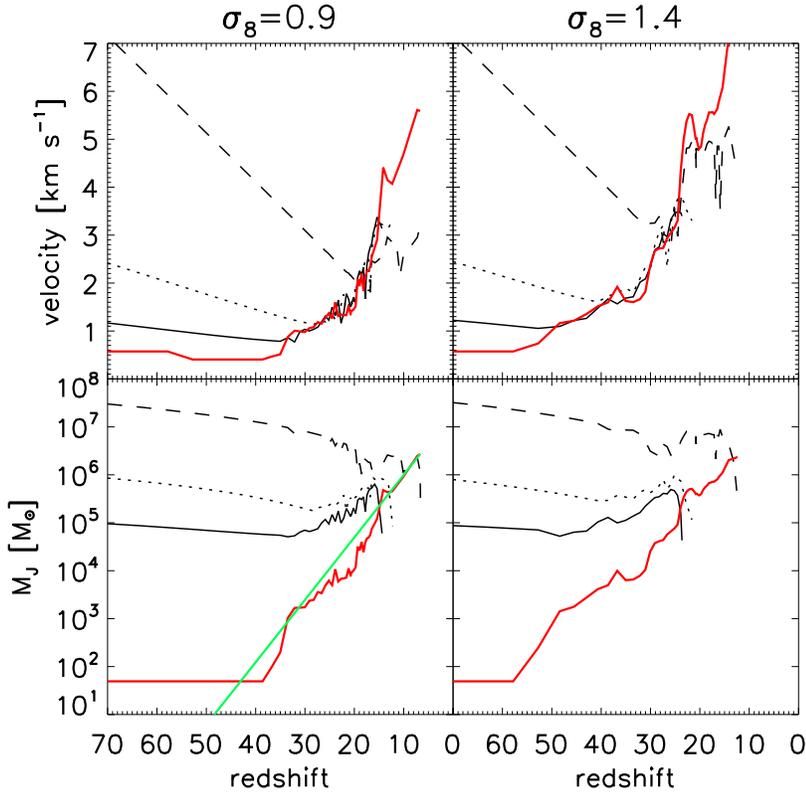}
\caption{{\it Top panels}: Effective velocity $\rm v_{\rm eff} =
\sqrt{{\it c}_{\rm s}^2 + \rm v_{\rm s}^2}$ of the gas (thin lines) and virial
velocity $V_{\rm vir}$ of the simulated minihalo (thick red line).
{\it Top Left}: `Standard collapse' case.  Dashed line: $\rm v_{\rm s,i} = 10$ km s$^{-1}$, 
dotted line: $\rm v_{\rm s,i} = 3$ km
s$^{-1}$, solid black line: no-streaming case. At each
redshift $\rm v_{\rm s}$ was found by taking an average over the entire
gas within the simulation box.  $c_{\rm s}$ refers to the average
sound speed of all particles within the virial radius of the minihalo.
{\it Top Right}: Early collapse case. Note that for the streaming
cases, the redshift at which $\rm v_{\rm eff}$ first falls below $\rm v_{\rm vir}$ 
matches well with the point where the gas thermal evolution
first follows that of $\rm V_{\rm vir}$.  {\it Bottom Panels}: Evolution
of the Jeans mass $M_{\rm J}$ with redshift, evaluated using $\rm v_{\rm eff}$ 
in the role of the effective sound speed. Notation is the same
as in the upper panels.  Red line now shows virial mass $M_{\rm vir}$
of the minihalo.  Green line is an exponential fit to the growth of
the `standard collapse' case minihalo.  Gas collapse occurs quickly
after $M_{\rm J}$ drops below $M_{\rm vir}$.  Note that the
enhancement of $\rm v_{\rm eff}$ due to the streaming velocity effectively
increases $M_{\rm J}$, causing the gas collapse to be delayed until
$M_{\rm vir}$ can further grow. This alters the final gas collapse
redshifts of each case ($z_{\rm col} = 14.4,\, 12.2,$ and $6.6$ for
the no streaming, moderate streaming, and fast streaming cases given
`standard collapse'; $z_{\rm col} = 23.6, \,21.3,$ and $12.4$ for `early
collapse'). }
\label{vel_vs_z}
\end{figure*}

\section{Numerical Methodology}

We carry out our investigation using GADGET, a widely-tested three
dimensional smoothed particle hydrodynamics (SPH) code
(\citealt{springeletal2001,springel&hernquist2002}). Simulations are
performed in a periodic box with size of 100~$h^{-1}$~kpc (comoving)
and initialized at $z_{\rm i}=100$ with both DM and SPH gas particles.  This
is done in accordance with a $\Lambda$CDM cosmology with
$\Omega_{\Lambda}=0.7$, $\Omega_{\rm m}=0.3$, $\Omega_{\rm B}=0.04$,
and $h=0.7$.  We adopt $\sigma_8=0.9$ for the fiducial normalization
of the power spectrum, and also examine the case of $\sigma_8=1.4$ in
which structure formation is accelerated and the first minihalo
collapses earlier.  Each simulation box contains $128^3$ DM particles
and an equal number of SPH particles.  The gas particles each have a
mass $m_{\rm SPH} = 8$~M$_{\odot}$, so that the mass resolution is,
$M_{\rm res}\simeq 1.5 N_{\rm neigh} m_{\rm SPH}\la 400$ M$_{\odot}$,
where $N_{\rm neigh}\simeq 32$ is the typical number of particles in
the SPH smoothing kernel (e.g. \citealt{bate&burkert1997}).  This mass
resolution allows us to follow the gas evolution to a maximum number
density of $n_{\rm max} = 10^{4}$ cm$^{-3}$.

The chemistry, heating and cooling of the primordial gas is treated in
a fashion very similar to previous studies
(e.g. \citealt{bromm&loeb2004,yoshidaetal2006}).  We follow the
abundance evolution of H, H$^+$, H$^-$, H$_2$, H$_{2}^{+}$, He,
He$^+$, He$^{++}$, e$^-$, and the deuterium species D, D$^+$, D$^-$,
HD, and HD$^+$.  We use the same chemical network as used in
\citet{greifetal2010} and include the same cooling terms.

We first perform both the `standard collapse' ($\sigma_8=0.9$) and
`early collapse' ($\sigma_8=1.4$) initializations with no streaming
velocity added.  For each of these we also perform `moderate' and
`fast' streaming cases in which we include an initial streaming
velocity $\rm v_{\rm s,i}$ of 3 km s$^{-1}$ and 10 km s$^{-1}$,
respectively.  The `moderate' streaming case represents the
 predicted root mean square velocity (\citealt{tse&hirata2010}), given that peculiar velocities have decreased as $(1+z)$ since recombination and thus have declined by a factor of 10 at the point our simulations are initialized. Our $\rm v_{\rm s,i}$ values therefore correspond to velocities of 30 km s$^{-1}$ and 100 km s$^{-1}$ at recombination, similar to the velocities chosen by \cite{maioetal2010}, 30 and 60 km s$^{-1}$.

\begin{figure}
\includegraphics[width=0.45\textwidth]{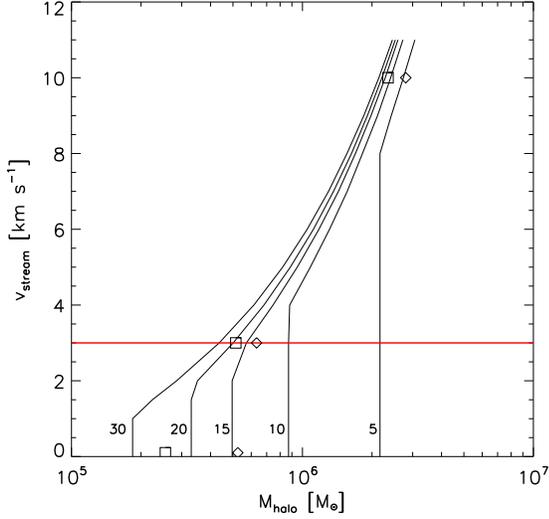}
\caption{Effect of relative streaming on the minimum halo mass into which primordial gas can collapse.  Each line represents the necessary halo masses for baryon collapse at a different redshift, marked in the plot.  The diamonds represent the final halo masses found in  `standard collapse' simulations ($z_{\rm col} = 14$ for no streaming), and the squares represent masses from the `early collapse simulations' ($z_{\rm col} = 24$ for no streaming).  Note that the halo mass does not noticeably increase unless the initial streaming velocites are very high ($\ga 3$  km s$^{-1}$). Also note that halos collapsing at high redshift are more affected by relative streaming, as the physical streaming velocities are higher at these early times. }
\label{vs_vs_m}
\end{figure}

\section{Results}

\subsection{Delay of Gas Collapse}

The main effect of the relative streaming cases is to delay the
collapse of the baryons into the DM halos.  In the standard case, the
collapse redshifts are $z_{\rm col} = 14.4,\, 12.2,$ and $6.6$ for
 $\rm v_{\rm s,i} =$ 0, 3, and 10 km s$^{-1}$  (0, 30, and 100 km s$^{-1}$ at recombination).  
The streaming
cases correspond to delays in collapse of $\sim 7 \times 10^{7}$ and
$\sim 5 \times 10^8$ years, respectively.  In the accelerated collapse
case these values are $z_{\rm col} = 23.6, \,21.3,$ and $12.4$,
corresponding to delays of $\sim 2 \times 10^{7}$ and $\sim 2 \times
10^8$ years.  Thus, this delay is noticeable only for high initial
values of $v_{\rm s,i} \ga 3$ km s$^{-1}$, whereas at smaller
values the delay is small compared to the Hubble time.

We can understand the criterion for gas collapse in terms of the
cosmological Jeans mass.  In the usual no-streaming case, the slow
infall of gas into the halos will first begin when the gravitational
potential well of the minihalo, characterized by its virial velocity
$V_{\rm vir}$, is large enough to assemble the gas, which occurs when
$V_{\rm vir} > c_{\rm s}$, where
$c_{\rm s} = \sqrt{k_B T/\mu m_{\rm H}}$ is the sound speed.  
Once this process begins, the sound speed
$c_{\rm s}$ will be coupled to $V_{\rm vir}$ through adiabatic heating
(see top panels of Fig.~\ref{vel_vs_z}), and the density will scale
with sound speed approximately as $c_{\rm s}^{3}$.  In
Fig.~\ref{vel_vs_z}, we determined the properties of the largest halo
in our simulation using the HOP technique
(\citealt{eisenstein&hut1998}) to find the DM particle in the region
of highest DM density.  Assuming this particle marks the center of the
halo, the extent of the halo was determined by finding the surrounding
spherical region in which the average DM density is 200$\rho_{c}$,
where $\rho_{c}$ is the redshift-dependent critical density.

The bottom panels of Fig.~\ref{vel_vs_z} illustrate that the adiabatic phase of evolution will continue until the virial mass of the minihalo is greater than the Jeans mass of the gas, $M_{\rm vir} > M_{\rm J}$.  For the no-streaming case, we calculate $M_{\rm J}$ as

\begin{equation}
M_{\rm J} = \left( \frac{\pi}{6}\right) \frac{c_{\rm s}^3}{G^{3/2}\rho^{1/2}}  \mbox{\ ,}
\end{equation}

\noindent Once the halo gains sufficient mass, and also provided that
the H$_2$-driven cooling time $t_{\rm cool}$ of the gas is shorter
than its free-fall time $t_{\rm ff}$, the Jeans and cooling criteria
will be satisfied and the gas will begin the next phase of rapid
collapse to higher densities, quickly reaching $n_{\rm max} = 10^{4}$
cm$^{-3}$.

The cause for the delay in collapse of the streaming cases lies in the enhanced effective velocity of the gas,

\begin{equation}
{\rm v}_{\rm eff} = \sqrt{{\it c}_{\rm s}^2 + {\rm v}_{\rm s}^2}  \mbox{\ ,}    
\end{equation}

\noindent where the streaming velocity decreases with redshift as
${\rm v}_{\rm s}(z)~=~{\rm v}_{\rm s,i}/(1 + z)$. As shown in the top panels of
Fig.~\ref{vel_vs_z}, this delays the point at which the gas will begin
falling into the halo.  To accommodate the cases with streaming, we
replace $c_{\rm s}$ in equation~(1) with $\rm v_{\rm eff}$, and the resulting
increase of $M_{\rm J}$ is shown in the bottom panels of
Fig.~\ref{vel_vs_z}.

For any given collapse redshift, a larger $M_{\rm vir}$ is therefore
required to trigger the collapse of streaming gas compared to the
non-streaming case.  In Fig.~\ref{vs_vs_m}, we estimate for different
redshifts the mininum halo mass into which gas with various initial
streaming velocities can collapse.  We arrive at these estimates using
the following simple prescription. We first determine for a range of
redshifts the minihalo mass corresponding to a virial temperature of
1500 K, which serves as the minimum mass for collapse and cooling
given no streaming.  We fit a typical halo growth history using
$M_{\rm vir}(z)~=~M_0 e^{\alpha z}$ (the green line shown in
Fig.~\ref{vel_vs_z}), with $M_0 = 2\times 10^7$~M$_{\odot}$ and
$\alpha$ ranging from $-0.2$ to $-0.5$.  We vary $\alpha$ depending on
the no-streaming case minihalo mass and the desired collapse redshift.
For every given collapse redshift and streaming velocity, we first
determine the redshift $z_{\rm eq}$ where $V_{\rm vir}(z) = \rm v_{\rm eff}(z)$.  
We assume that $z_{\rm eq}$ is the point where the gas
switches from having properties of the intergalactic medium (IGM) to
properties determined by the halo.  Thus, for $z > z_{\rm eq}$ the
sound speed roughly follows that of the IGM, $c_{\rm s, IGM}$.
Therefore, $z_{\rm eq}$ can be found by considering
\begin{equation}
V_{\rm vir}(z) = \sqrt{G\,M_{\rm vir}(z)/R_{\rm vir}(z)} = \sqrt{c_{\rm s, IGM}(z)^2 + \rm v_{\rm s}(z)^2}  \mbox{\ ,}    
\end{equation}
where 
\begin{equation}
R_{\rm vir}(z) \simeq 210 \left(\frac{M_{\rm vir}}{10^6 \rm M_{\odot}} \right)^{1/3} \left(\frac{1+z}{10} \right)^{-1}f(z)\, \mbox{\,pc,}
\end{equation}
and $f(z)$ is a factor of order unity with a mild dependence on
redshift (\citealt{barkana&loeb2001}).  At $z = z_{\rm eq}$
the effective gas velocity is thus ${\rm v}_{\rm eq} = V_{\rm vir}(z_{\rm eq})$.  After this point the thermal energy of the halo gas dominates over the energy of streaming motion, and its sound speed can be described by the halo virial velocity thereafter.  

\begin{figure*}
\includegraphics[width=.4\textwidth]{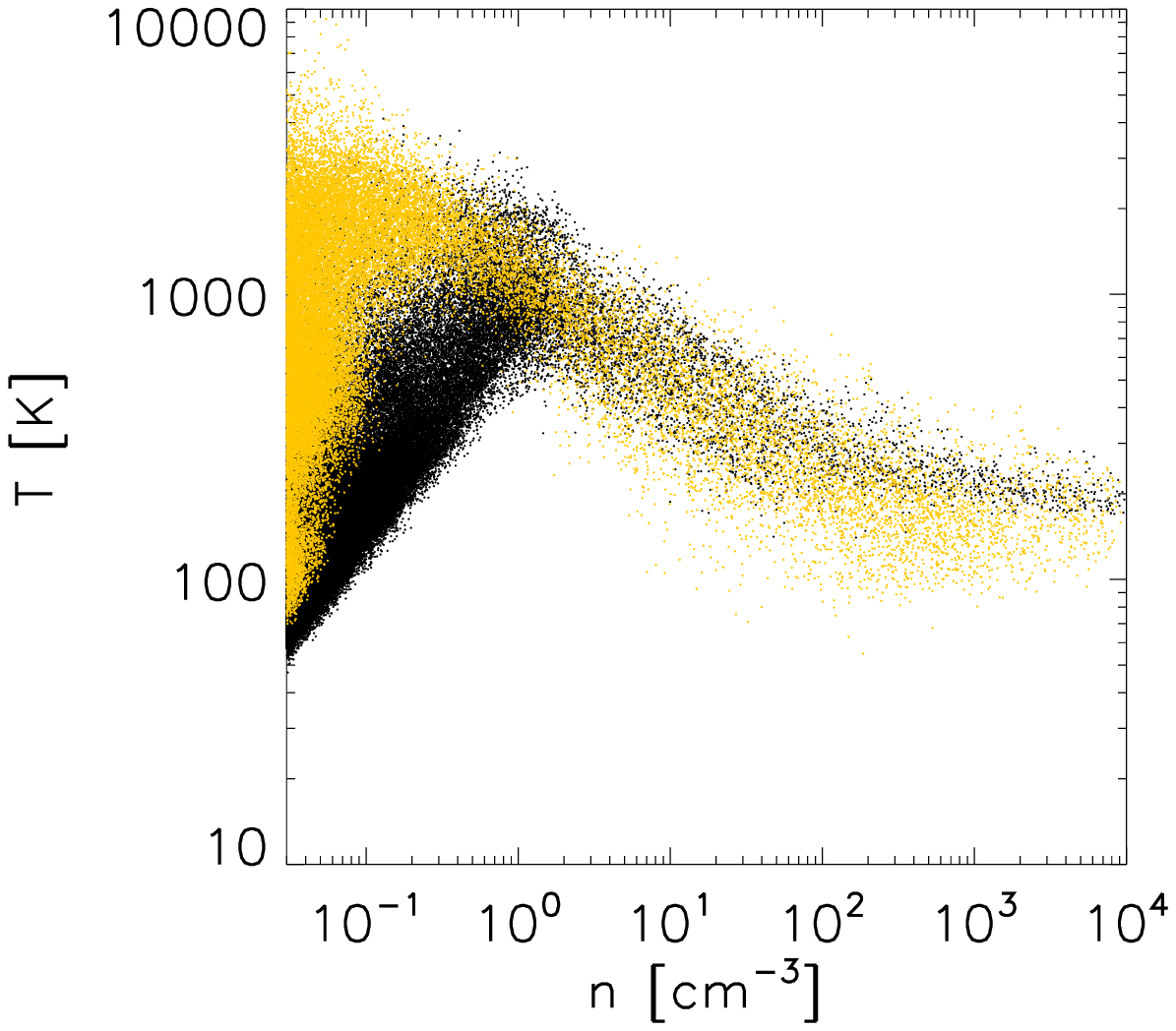}
\includegraphics[width=.4\textwidth]{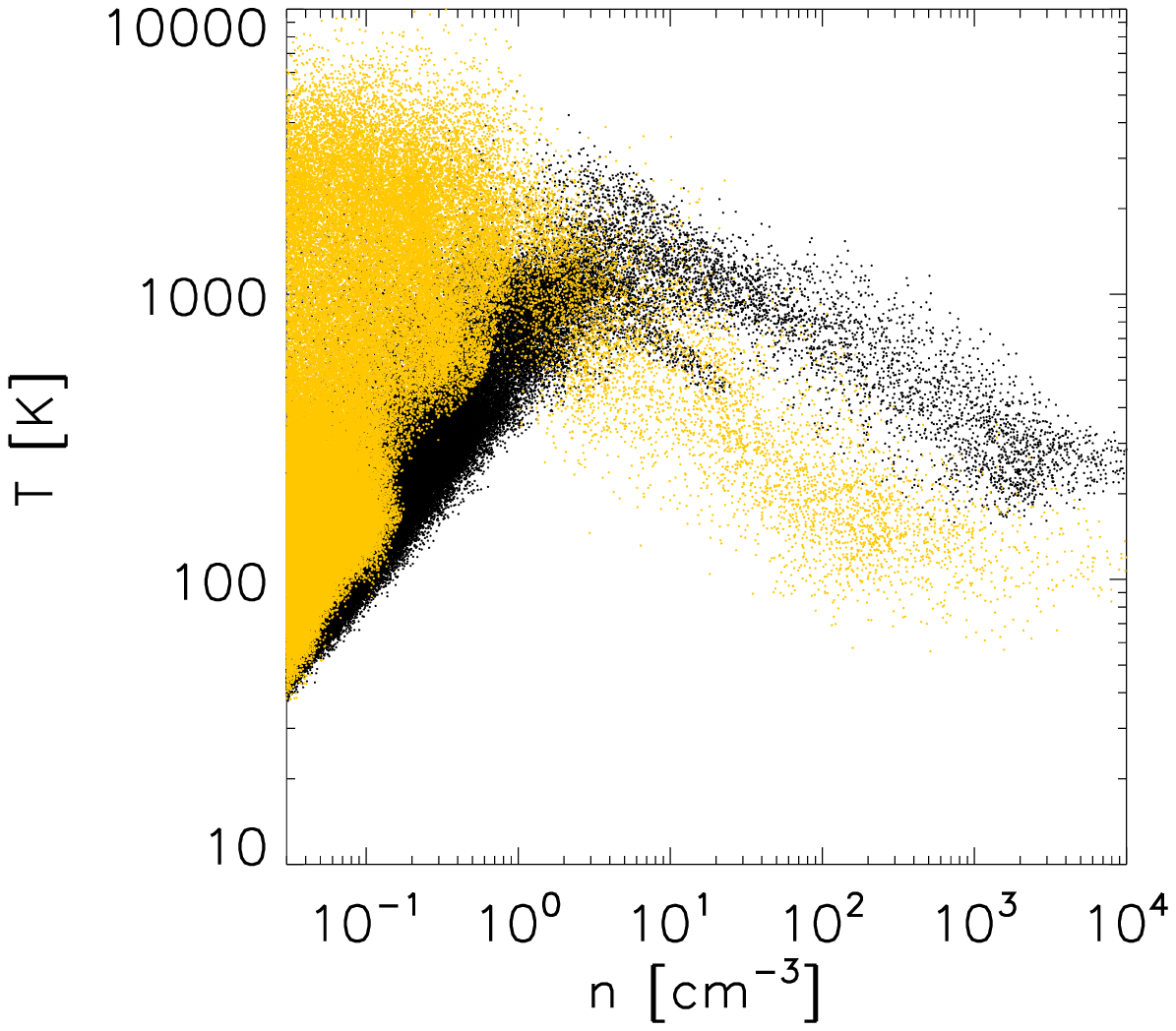}
\caption{ Evolution of temperature of the gas as density grows for both `standard' and `early' collapse cases.
Black dots: No streaming case.  Yellow dots: $\rm v_{\rm s,i} = 10$ km s$^{-1}$ case.
 {\it Left}:  Standard case is shown at $z=14.4$, 
while the $\rm v_{\rm s,i}$ case is shown at $z=6.6$.  
{\it Right}:  Early collapse case is shown at $z=23.6$ for no streaming, and $z=12.4$ for $\rm v_{\rm s,i} = 10$ km s$^{-1}$.
There is almost no difference
between the $\rm v_{\rm s,i} = 3$ km s$^{-1}$ (not shown) and no-streaming
cases.}
\label{bf10}
\end{figure*}

Furthermore, $z_{\rm eq}$ marks the last time that the gas density within $R_{\rm vir}$ is still that of the IGM.  
The density of halo gas when it first couples to the DM is then
\begin{equation}
\rho_{\rm eq} = \rho_{\rm IGM}(z_{\rm eq}) \simeq 2 \times 10^{-29}
\Omega_{\rm m} h^2 (1 + z_{\rm eq})^3 ~{\rm g~cm^{-3}}.
\end{equation}
Note that, because $z_{\rm eq}$ is lower for higher values of $\rm v_{\rm s, i}$, 
\,$\rho_{\rm eq}$ correspondingly decreases.

Finally, as the gas infall into the halo continues for $z < z_{\rm eq}$, we estimate its average density to be 
\begin{equation}
\rho(z) \simeq \rho_{\rm eq}\left(\frac{V_{\rm vir}(z)}{\rm v_{\rm eq}}\right)^{3} \mbox{\ .} 
\end{equation}
The above equation describes how the gas density will adiabatically evolve with thermal energy as it collapses (e.g. \citealt{tegmarketal1997}).  Recall that $\rm v_{\rm eq}$ is the effective sound speed when it first begins falling into the halo, and that adiabatic evolution implies $T \propto \rho^{\gamma - 1} = \rho^{2/3}$ for an atomic gas with $\gamma=5/3$. Using $c_{\rm s} \propto T^{1/2}$ results in $\rho \propto c_{\rm s}^3$. Finally, we replace $c_{\rm s}$ with the virial velocity of the halo to arrive at the approximation in Equ.~(6). 
The density will increase in this way until the gas virializes and reaches a maximum of $200\rho_{\rm IGM}(z)$.  Inserting the applicable values for $\rho$
and $V_{\rm vir}$ at the given collapse redshift, $z_{\rm col}$, we
arrive at $M_{\rm J}(z_{\rm col})$.

This model well reproduces the masses and collapse redshifts found in
the simulations (symbols in Fig.~\ref{vs_vs_m}). As the streaming
velocities increase, the gas density during initial infall decreases,
thereby  lowering the typical gas density in the halo and raising the minimum
mass that will satisfy $M_{\rm vir} > M_{\rm J}$.  For the average 3
km s$^{-1}$ streaming velocities, this minimum mass $M_{\rm halo}$
will approximately double for $z_{\rm col} = 30$, but will almost stay
the same by $z_{\rm col} = 10$.

\begin{figure*}
\includegraphics[width=.7\textwidth]{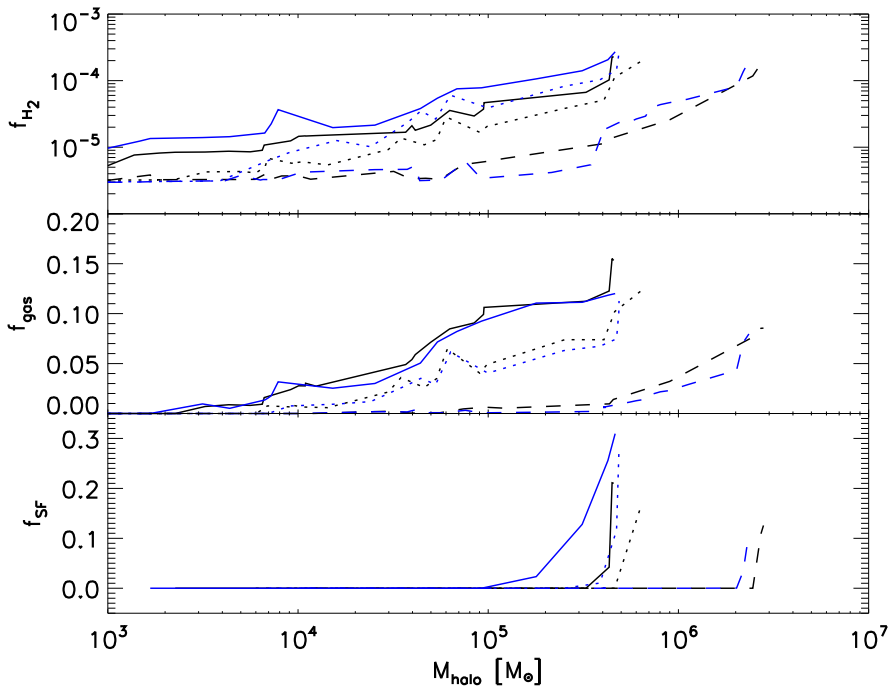}
\caption{ Evolution of gas properties with halo mass.  Dashed lines: $\rm v_{\rm s,i} = 10$ km s$^{-1}$,
dotted lines: $\rm v_{\rm s,i} = 3$ km s$^{-1}$, 
solid lines: no streaming.  Black represents the `standard collapse' set of simulations, while blue represents the `early collapse' set.  {\it Top panel}: Average H$_2$ fraction of the minihalo gas. {\it Middle panel}:  Gas fraction f$_{\rm gas}$ of the halos. {\it Bottom panel}: Fraction of minihalo gas that is star-forming (i.e., $n >$ 1 cm$^{-3}$).  The reduced gas fraction and the delay of star formation for high streaming velocities are evident in the bottom panels.  However, the H$_2$ fraction converges to $\ga 10^{-4}$ in each case, allowing for the thermal evolution of the highest density gas to be relatively unchanged even for the highest streaming velocites.}
\label{sf}
\end{figure*}

\subsection{Thermal Evolution and Star Formation}

Figure~\ref{bf10} compares the thermal evolution for the  `standard' and
`early' collapse cases given no streaming with the corresponding rapid 
streaming cases ($v_{\rm s,i} = 10$
km s$^{-1}$).  For the no streaming cases, the gas follows the
canonical evolution of adiabatic heating as the IGM gas gradually
becomes incorporated into the growing minihalo.  This gas heats to the
virial temperature ($\sim 1000$ K) of the minihalo until the H$_2$
fraction grows sufficiently high to allow the gas to cool to a minimum
of $\simeq 200$ K.  As the gas temperature drops, its density grows to
approximately $10^4$ cm$^{-3}$ (see
\citealt{brommetal2002,yoshidaetal2006}).  At this density the gas has
reached the `loitering phase,' and this is the reservoir of gas from
which Pop III stars will form.

For the $\rm v_{\rm s,i} = 10$ km s$^{-1}$ case, this evolution shows only
minor differences from that described above.  The initial heating of
the low-density gas occurs more quickly than the purely adiabatic
rate, and the streaming velocity acts as a heating term for the
low-density gas.  However, once the gas gains sufficiently high
temperature and H$_2$ fraction, the gas cools and condenses to
approximately 200 K and $10^4$ cm$^{-3}$, just as in the canonical
case, though the minimum temperature is slightly lowered 
for the streaming cases.  
Subsequent star formation is therefore not suppressed, and
should occur in the same way as it would in the no streaming case.
Note also that Fig.~\ref{bf10} represents significantly higher
streaming velocities than typically expected.  For the more representative $\rm v_{\rm s,i} =
3$ km s$^{-1}$ cases, the thermal evolution shows almost no difference
from those with no streaming.  This further strengthens the argument that
relative streaming between baryons and dark matter will do little to
modify Pop III star formation.

 Figure~\ref{sf} further elucidates the effect of relative streaming on gas collapse
and star formation. The delay of gas collapse until the minihalos have reached higher masses is evident in the bottom panel, 
which shows the fraction $f_{\rm SF}$ of the minihalo gas that is dense and star-forming, defined as the gas that exceeds densities of 1 cm$^{-3}$.  There is no star-forming gas in $< 10^6$M$_\odot$ halos for the highest streaming velocities.  The gas fraction $f_{\rm gas}$, calculated as the gas mass in the halo over its total mass, is reduced by up to a factor of $\sim 1.2$ for $\rm v_{\rm s,i} =$ 3 km s$^{-1}$, and a factor of $\sim$ 1.8 for $\rm v_{\rm s,i} =$ 10 km s$^{-1}$, even after the gas has reached high densities.  However, once gas collapse has occured, the average H$_2$ fraction $f_{\rm H_2}$ is very similar in all cases, as is the subsequent thermal evolution (Figure~\ref{bf10}).

\section{Discussion}

Our series of simulations show that Pop III star formation will be
essentially the same in cosmologies with relative streaming motions
between gas and DM, even in regions with streaming velocities much
higher than average ($\rm v_{\rm s,i} \ga 3$ km s$^{-1}$).  However, these
regions of fast streaming will experience a modest delay in the
collapse redshift at which Pop III stars will first form, while in
regions of typical streaming the delay will be minimal ($\ga 10^7$
years), in good agreement with \cite{maioetal2010}.   In their work they also find similar reductions in halo gas fractions of up to a factor of 2, even given their larger box size and lower resolution. This is furthermore consistent with other recent work such as that of \cite{tseetal2010}.

The effect on reionization should be similarly minimal.  Though not
yet known with certainty, recent work has suggested that the sources
of reionization were dominated by early galaxies of virial
temperatures above the hydrogen cooling threshold of $10^4$K
(corresponding to masses $\gtrsim 10^8$M$_\odot$), with a much smaller
contribution from $\lesssim 10^6$M$_\odot$ halos
(e.g. \citealt{trac&gnedin2009,trenti&stiavelli2009,munoz&loeb2010}).  
The relative streaming motions will do little to alter the infall of
gas into the larger potential wells of ionizing galaxies, and thus
reionization should proceed virtually unaffected. The effect of
streaming is mostly pronounced at the highest redshifts when the
collapse fraction and the corresponding radiative effects of stars are
exceedingly small.

In conclusion, we have directly simulated the delay in collapse, and
the subsequent thermal evolution of the first baryonic structures
under relative bulk velocities between gas and dark matter.  Our
results show that early star formation and subsequent evolution of
reionization should quickly converge to the no-streaming case.  Thus,
results of previous and future cosmological studies concerning the
formation of the first stars and galaxies will need only minimal
modifications due to the relative streaming effect.

\acknowledgments

VB acknowledges support from NSF grants AST-0708795 and AST-1009928, as well as NASA ATFP grants NNX08AL43G
and NNX09AJ33G. The simulations were carried out at the Texas Advanced
Computing Center (TACC).

\bibliographystyle{apj}
\bibliography{bf}

\clearpage

\end{document}